\documentclass[%
 reprint,
 amsmath,amssymb,
 aps,
]{revtex4-1}

\usepackage{graphicx}
\usepackage{hyperref}
\hypersetup{
	colorlinks   = true,
	citecolor    = blue,
	linkcolor = blue,
    urlcolor = blue
}
\usepackage[inline]{enumitem}

\usepackage{float}
\usepackage{xcolor}
\usepackage{multirow}

\begin{document}


\title{Structural, Electronic, and Magnetic Properties of Bulk and Epitaxial LaCoO$_3$ through Diffusion Monte Carlo}


\author{Kayahan Saritas}
\email[]{saritask@ornl.gov}
\author{Jaron T. Krogel}
\author{Satoshi Okamoto}
\author{Ho Nyung Lee}
\author{Fernando A. Reboredo}
\email[]{reboredofa@ornl.gov}
\affiliation{Materials Science and Technology Division, Oak Ridge National Laboratory, Oak Ridge, Tennessee 37831, USA}


\date{\today}

\begin{abstract}
Magnetism in lanthanum cobaltite (LCO, LaCoO$_3$) appears to be strongly dependent on strain, defects, and nanostructuring. LCO on strontium titanate (STO, SrTiO$_3$) is a ferromagnet with an interesting strain relaxation mechanism that yields a lattice modulation. 
However, the driving force of the ferromagnetism is still controversial. Experiments debate between a vacancy-driven or strain-driven mechanism for epitaxial LCO's ferromagnetism. We found that a weak lateral modulation of the superstructure is sufficient to promote ferromagnetism. Our research also showed that ferromagnetism appears under uniaxial compression and expansion. Although earlier experiments suggest that bulk LCO is nonmagnetic, our Diffusion Monte Carlo calculations found that magnetic phases have a lower energy ground state for bulk LCO.  This article discusses recent experiments indicating a more complicated picture for the bulk magnetism and closer agreement with our calculations. The role of defects are also discussed through excited-state calculations.
\end{abstract}

\pacs{}

\maketitle

\section{Introduction} 
The electronic structure of correlated systems can be heavily dependent on their geometry and external stimuli such as epitaxial strain, temperature, and impurities. Using lattice distortions, the balance between crystal field splitting and Hund's exchange can be manipulated to drive metal/insulator and ferromagnetic/antiferromagnetic (FM/AFM) transitions \cite{Dagotto2005, Tokura2000, Ahn2004}. Lanthanum cobaltite (LCO, LaCoO$_3$) is an example of a material that becomes a ferromagnet under epitaxial strain   \cite{Biskup2014, Mehta2015,Gazquez2011a, Kwon2014a, Choi2012, Qiao2015}. 
Ferromagnetism in epitaxial LCO is particularly interesting, as the bulk material was thought to be nonmagnetic \cite{Koehler1957}.
The ability to control the ferromagnetism in epitaxial LCO could yield novel basic properties and new technological applications. 
A combination of external factors has been found to simultaneously affect the atomic spin states of LCO \cite{Zobel2002, Klie2007a, Radaelli2002a, Senars-Rodrguez1995, Podlesnyak2006, Saitoh1997, Asai1998, Phelan2006, Zhou2007, Fuchs2007, Fuchs2009}. Therefore, having a clear understanding of the origin of the spin transition can be extremely challenging.  

Several authors \cite{Biskup2014, Mehta2015,Gazquez2011a, Kwon2014a, Choi2012, Qiao2015} have reported an unconventional mechanism of strain relaxation in epitaxial LCO. In scanning transmission electron microscope (STEM) images, LCO grown on strontium titanate (STO) typically yields a superstructure of two bright stripes following a dark stripe. In these images, brightness is associated with the larger electron density of La atoms. The brighter stripes indicate a smaller La-La separation ($\sim$3.61 {\AA}), whereas the La-La separation in the darker stripes is larger ($\sim$4.54 {\AA}) on the in-plane axis \cite{Biskup2014}. 
In LCO, Co has a nominal charge of 3+. Thus, Co$^{3+}$ can have three different atomic magnetic moments: high-spin (HS, $t^4_{2g}e^2_{g}$, $S$=2), intermediate-spin (IS, $t^5_{2g}e^1_{g}$, $S$=1) and low-spin (LS, $t^6_{2g}e^0_{g}$, $S$=0, nonmagnetic).
Dark stripes with a larger La-La distance might be ascribed to the HS state, due to the diminished crystal field splitting with the larger Co-O interatomic distances.
Therefore, the HS state of Co$^{3+}$ in darker stripes might explain the ferromagnetism observed in LCO thin films.
Two mechanisms are proposed to explain the lattice modulation and the ferromagnetism:
The first mechanism indicates that an ordered array of oxygen vacancies is the driving force for the superstructure formation and the ferromagnetism \cite{Biskup2014, Mehta2015,Gazquez2011a}.
However, the second mechanism indicates that the epitaxial strain drives the ferromagnetism through the rearrangement of the Co-octahedra  \cite{Kwon2014a, Choi2012, Qiao2015}. The controversy between the two experimentally suggested mechanisms arises from the interpretation of the methods, which provides averaged information. These mechanisms were also studied using Density Functional Theory (DFT) \cite{Kohn1965, Hohenberg1964} to help resolve the controversy. However, the accuracy of DFT has not yet been sufficient to differentiate between the two mechanisms \cite{Biskup2014, Kwon2014a}.

\begin{figure*}[t]
	\includegraphics[width=\textwidth]{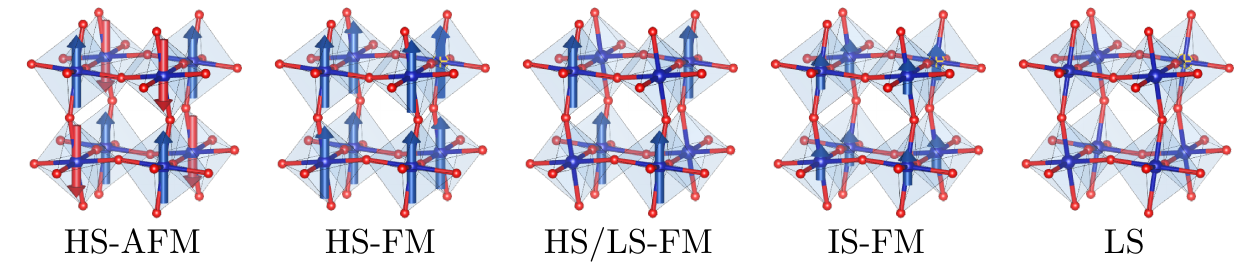}
	\caption{(Color online) LCO spin configurations investigated in this work. Cobalt atoms have high-spin (HS), intermediate-spin (IS), and low-spin (LS) states with ferromagnetic (FM) or antiferromagnetic (AFM) ordering. HS-AFM and HS-FM/LS orderings are G-type, meaning that spins are parallel aligned on (111), but this is not mentioned further on for brevity. Majority spin vectors are shown in blue, while minority spins are shown in red arrows. IS vectors are shown half the size of HS vectors. Co and O atoms are shown as blue and red circles, while La atoms are omitted for clarity.  \label{fgr:structures}}
\end{figure*}

Projector methods \cite{Ma2013}, such as Diffusion Monte Carlo (DMC) \cite{Ceperley1980, Foulkes2001}, are shown to be the most accurate and practical methods to tackle the ground states of complex, highly correlated materials with a similar success in the excited states \cite{Kolorenc2010, Yu2015, Needs2001, Mitra2015, Schiller2015, Saritas2018a, Saritas2017,Trail2016, Luo2016, Benali2016, Zheng2015,Kylanpaa2017, Yu2017,Santana2015, Shin2017, Shin2018, Saritas2019a}.  Although it is computationally more expensive, DMC explicitly accounts for the antisymmetry of the many-body wave function and electron correlation, without using any empirical parameters   \cite{Ceperley1980, Foulkes2001, Shulenburger2013, Needs2010}. DMC previously predicted the correct energetic ordering between the three polymorphs of CoO, surpassing the accuracy of DFT approximations \cite{Saritas2018b}. DMC was shown to yield accurate energies for La-containing compounds as well \cite{Santana2016a, Santana2017}. 
Thanks to its favorable computational scaling, $\mathcal{O}(N^3)$, DMC can be an ideal theoretical method to study LCO. 

In this work, we studied the magnetism of bulk and epitaxial LCO using DMC under isotropic scaling and lattice modulation. 
Understanding the magnetism of bulk LCO is an integral part of our study. Bulk LCO has long been thought to be nonmagnetic \cite{Koehler1957}, but recent experiments challenge this idea \cite{Belanger2016, Durand2015a, Durand2015, Kaminsky2018b, Durand2013,Belanger2016, Itoh1995}. 
We predict an AFM ground state for the bulk LCO using DMC. 
In uniaxially strained LCO, we find a complete transition from HS-AFM to HS-FM at a La-La separation of nearly 4.5 {\AA}.
When La-La separation is allowed to contract, another transition between HS-AFM and HS/LS-FM mixed phase is also observed at 3.71 {\AA}.
In epitaxial LCO, we find that a mixed phase of HS-AFM and HS/LS-FM phases is allowed to exist that is less than 0.1 eV per Co above the LCO ground state. 
Yet, the stable lattice modulation of the epitaxial LCO from DMC is not as large as the lattice modulation observed in the experiments.  
Our findings suggest that defects may be playing a more dominant role in driving this lattice modulation. 
We calculated the quasiparticle and optical gaps of LCO using various DFT approximations and DMC. Our results indicate LCO optical and quasiparticle gaps of 3.75 eV from DMC, while 1.6--1.9 eV Kohn-Sham gaps from DFT. Our theoretical estimates of these gaps are significantly larger than the experimental estimates, which are less than 1 eV. We argue that the presence of defects in LCO or internal $d$-$d$ transitions provides a possible explanation for both the substantially large lattice modulation and the overestimation of the experimental gap energies. 

Section \ref{sc:method} briefly describes the DMC approach and our methods. In Section \ref{sec:eos}, using the experimental geometry from the literature, we study the ground-state magnetic and structural properties of bulk LCO using DMC. In Section \ref{sec:latmod}, we first calculate the ground-state DMC energies of various magnetic phases in epitaxial LCO with uniform La-La separation. Then, under the epitaxial equilibrium conditions, we study the lattice modulation (i.e., superstructure formation). 
Section \ref{sec:latmod} shows that external factors such as defects are required to drive larger lattice modulations. 
In Section \ref{sec:excited}, we present the results of our DFT and DMC calculations for optical and electronic gaps, which suggest the existence of intrinsic defects. 
In Section \ref{sec:oo}, we study the orbital ordering in LCO, which provides the physical reasoning for HS-FM ordering over HS-AFM as observed in Section \ref{sec:latmod}. In Section \ref{sec:dft}, we present our DFT and DMC benchmark for the magnetic ground states of bulk LCO studied in Section \ref{sec:eos}. Finally, in Section \ref{sec:conc}, we provide study conclusions and discuss future research. 

\section{Methods}\label{sc:method}
For this work, we used DMC to obtain ground-state energies of bulk and epitaxial LCO, as well as the excited-state energies of bulk LCO. Methods for calculating excited states will be discussed later in this section. DFT functionals used in this work include local (LDA \cite{Perdew1981}), semilocal (PBE \cite{Perdew1996} and PBEsol \cite{Perdew2008}), and meta-GGA (SCAN \cite{Perdew2017}) functionals involving benchmarks with hybrid-DFT (B3LYP \cite{Becke1993b}, HSE \cite{Heyd2003}, and PBE0 \cite{Adamo1999a}) from the literature. In addition, Dudarev's Hubbard-$U$ \cite{Anisimov1997, Dudarev1998} corrected LDA, PBE, and PBEsol functionals are also used to avoid well-known self-interaction error in correlated systems. 

Geometry relaxation in DMC is possible for extended systems, but it is still computationally intensive and challenging \cite{Archibald2018, Sorella2010}. Due to this challenge, DMC studies often use experimental structural parameters with no relaxation \cite{Archibald2018}. Similarly for bulk LCO, experimental structural parameters are available from the neutron diffraction experiments done at 4 K (ICSD no. 201761) \cite{Thornton1986}. However, for epitaxial LCO, the experimental structures are not available. Therefore, a structural optimization method that is empirically validated and can best reproduce the experimental structures can be used in DMC. Ref. \citenum{Belanger2016} shows that PBEsol+$U$ = 4 eV is superior to PBE+$U$ and LDA+$U$ functionals in terms of producing accurate Co-O-Co bond angles and the equilibrium volume of bulk LCO. Therefore, we used PBEsol+$U$ = 4 eV to obtain the structural parameters of epitaxial LCO to be used in DMC calculations. For the DMC calculations on bulk LCO, the experimental structural parameters are used with no relaxation. For the DFT benchmark calculations, the geometry optimization is performed separately with each functional. Benchmark DFT calculations and band decomposed charge density calculations were performed using \textsc{VASP} code with projector augmented wave (PAW) pseudopotentials \cite{Kresse1996, Kresse1999}, using a kinetic energy cutoff of 520 eV with 6x6x6 reciprocal grid.  

The Variational Monte Carlo (VMC) and DMC \cite{Foulkes2001, Needs2010} calculations were performed using \textsc{QMCPACK}   \cite{QMCPACK}, while DFT-VMC-DMC calculation workflows were generated using Nexus   \cite{Nexus} software suite. We used DMC trial wavefunctions in the Slater-Jastrow form \cite{Slater1929, Jastrow1955}. 
\textsc{Quantum Espresso} \cite{QE-2009} (QE) code was used to generate the single determinant spin-up and spin-down orbitals. We included terms up to three-body Jastrow correlation functions as described in \cite{Drummond2004f}. 
These functions were parameterized in terms of radial blip-splines for one- and two-body terms and in terms of low-order polynomials for the three-body terms. 
The purpose of using a trial-wavefunction with the Slater-Jastrow form is to guide the simulation to achieve the ground-state energy with higher accuracy, smaller localization error \cite{Mitas1991}, and reduced variance \cite{Foulkes2001}. Jastrow parameters are optimized using subsequent VMC variance and energy minimization calculations using the linear method \cite{Umrigar2007}. Cost function of the energy minimization is split as 95/5 energy and variance minimization, which is shown to provide a good balance for improvements in DMC with the resulting variance \cite{Umrigar2005}. 
The Slater part of the trial wavefunction is optimized by improving the nodal surface. DMC has the zero-variance property, meaning that as the trial wavefunction approaches the exact ground state (i.e., having exact nodal surface), statistical fluctuations in the energy reduce to zero \cite{Foulkes2001}. 
Various sophisticated methods can be used to optimize the nodal surface of the trial wavefunction  \cite{Kwon1993,Kwon1998,LopezRios2006, Bajdich2010}. 
However, we used a simpler approach with LDA+$U$, where the Hubbard-$U$ value is used as a variational parameter to optimize the nodal surface using DMC.
In the appendix, we show that DMC minima is largely insensitive to the choice of DFT functional, and the choice of $U$ value does not affect the ordering between the magnetic phases of LCO.
Our findings are also supported by previous studies on NiO \cite{Shin2018}, TiO$_2$ \cite{Luo2016}, and CoO \cite{Saritas2018b}. We found that LDA+$U$ = 6 eV gives optimized DMC energies for all the magnetic phases of LCO studied here; hence it is used throughout this work.

A timestep of 0.01 Ha$^{-1}$ and a supercell 3x3x3 reciprocal twist were used in all DMC calculations. Convergence tests regarding the timestep error and the one-body finite-size effects are given in the appendix and the supplementary information \cite{supp}.
The locality approximation is used to evaluate the nonlocal pseudopotentials within DMC \cite{Mitas1991}.
Compared to the T-moves approximation \cite{Casula2010}, the locality approximation is found to reduce the localization error further for the pseudopotentials used in this work \cite{Dzubak2017a, Krogel2017a}. 
Model periodic Coulomb (MPC)   \cite{Williamson1997, Kent1999} interaction was used to eliminate spurious two-body interactions on the potential energy   \cite{Drummond2008, Fraser1996}. 
We used hard LDA \cite{Perdew1981} RRKJ pseudopotentials that had been generated using OPIUM  \cite{OPIUM} and were previously tested for use in Quantum Monte Carlo (QMC) operations \cite{Krogel2016, Santana2016a, Santana2017, Saritas2018b}. 
The kinetic energy cutoff we used, 350 Ry, is found to converge total energies within 1 meV per atom. 

QMC simulations were performed using supercells containing a minimum of 8 formula units (40 atoms for bulk LCO) when the energies of two structures with identical lattice parameters are compared. When the energy difference between two structures with different lattice parameters is calculated, finite-size extrapolations are performed using up to 90-atom cells \cite{supp}. The real-space blip-spline basis sets used in the finite-size extrapolation calculations can have very large memory requirements on computational nodes, making large simulations inaccessible. 
Therefore, a hybrid orbital scheme  \cite{Luo2018}, separating core and interatomic regions, was used in the finite-size extrapolation calculations. 
Within this scheme, cutoff radii of 2.2 {\AA}, 1.4 {\AA}, and 1.2 {\AA} were used on La, Co, and O atoms, respectively. The radii values we used for Co and O closely resemble the values used for Ni and O in NiO \cite{Luo2018}. 
These values were found to provide ground-state energies converged under 10 meV per atom \cite{supp}. 

\begin{figure}[t]
	\includegraphics[width=0.5\textwidth]{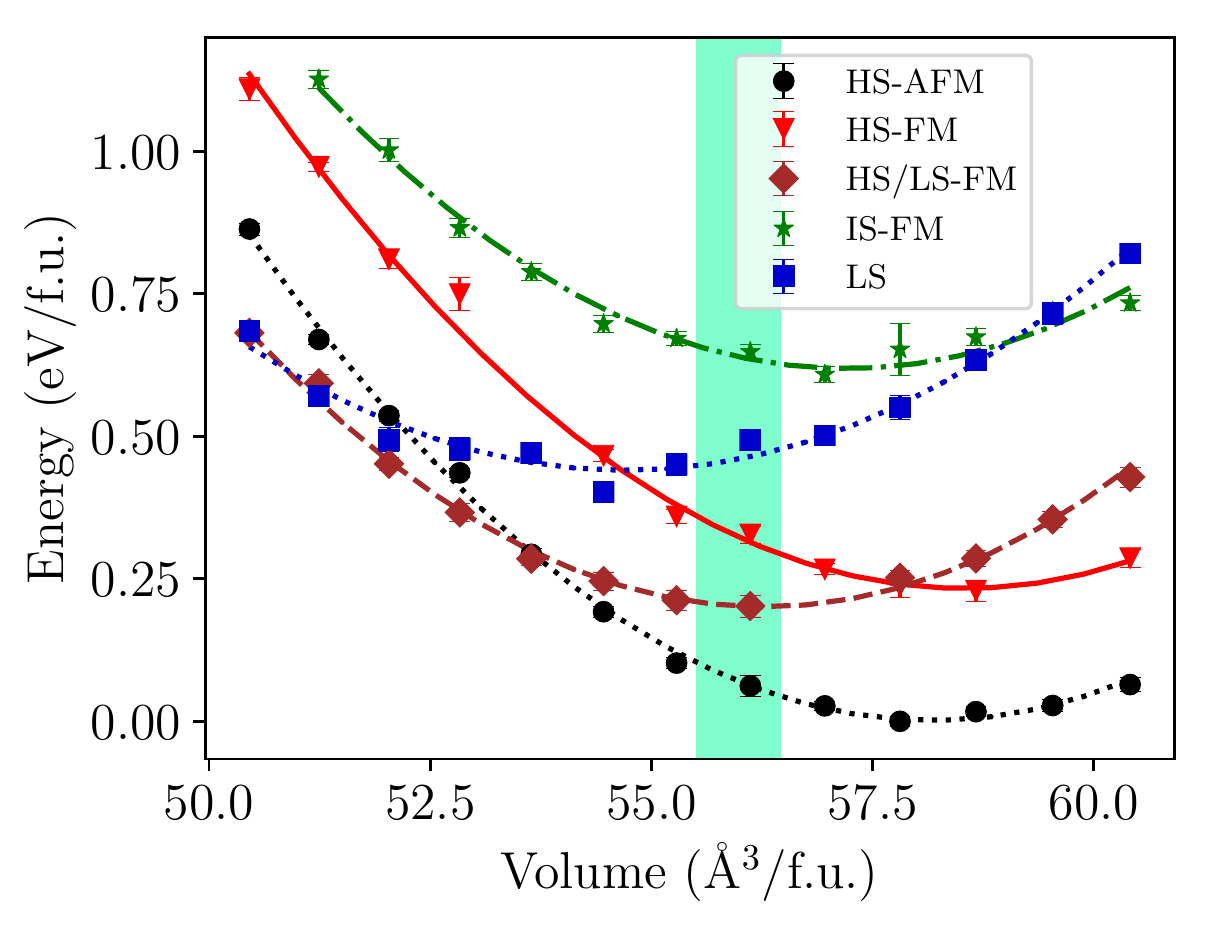}
	\caption{(Color online) DMC equation of states curves using isotropically scaled experimental structure \cite{Thornton1986}. Co-O-Co angle is set to 161.1 degrees. The minimum energy from the fits is used as the reference energy with -451.0164 Ha/f.u., using a 20-atom simulation cell. Green region shows the uncertainty in the experimental volume of LCO. \label{fgr:eos}}
\end{figure}

Excited-state calculations in DMC are done using automated workflows developed using \textsc{Nexus}. In this workflow, the primitive cell of the structures is standardized using \textsc{Spglib} \cite{Togo2018}, and the irreducible Brillouin zone paths are obtained using \textsc{SeeK-path} \cite{Hinuma2017} for the band structure calculations. Starting from an upper diagonal Hermite normal form  \cite{Lloyd2015}, tiling matrices are optimized for each twist for the largest Wigner-Seitz radius possible to reduce finite-size errors. 
While the experimental bulk LCO structure has a perfect Co octahedra, the geometry optimization yields a slight distortion, modifying the location of valence band maximum (VBM) and conduction band minimum (CBM). Following the geometry optimization with PBEsol+$U$ = 4 eV, LDA+$U$ = 6 eV calculations are used to determine VBM/CBM and to generate trial wavefunctions. Orbitals for the optical and quasiparticle calculations are extracted from the neutral ground-state wavefunction. An optical excitation is produced by annihilating the electron at VBM and creating another electron at CBM. To find the optical gap, the energy of this excited state, $E^{ex}$, is subtracted from the energy of the ground state, $E_N$, where $N$ is the number of electrons at the ground state. Therefore, the energy of the optical gap, $E_g$, is defined as $E_g = E^{ex}-E_N$. For the quasiparticle gap calculations, the ground-state energies of positively and negatively charged cells, $E_{N+1}$ and $E_{N-1}$, respectively, are used. The quasiparticle gap, $E^{QP}$, is defined as $E^{QP} = E_{N+1} + E_{N-1} - 2E_{N}$, which is equal to $E^{QP} = E_a-IP$, such that $E_a$ is the electron affinity and $IP$ is the ionization potential.

\section{Results and Discussion}
\subsection{Isotropic scaling of bulk LCO}\label{sec:eos}

Bulk LCO has a rhombohedral, \textit{R$\bar{3}$c} structure ($a=5.35$ {\AA}, $\beta=60.96^\circ$), with the Co$^{3+}$ ions having $d^6$ valence structure   \cite{Thornton1986, Autret2005,Øygarden2012}. 
The ground spin state of Co$^{3+}$ in bulk LCO (T $<$ 30 K) is reported to be LS \cite{Wollan1955, Koehler1957, Heikes1964, Raccah1967, Raccah1967, Goodenough1958a, Goodenough1967}, though current experiments now question this long-held conclusion. In the low-temperature region, AFM correlations are shown to dominate the coexisting FM correlations \cite{Belanger2016, Durand2015a, Durand2015, Kaminsky2018b, Durand2013,Belanger2016, Itoh1995}. This is different from an LS scenario where all Co atoms would be diamagnetic. 
However, an AFM ground state cannot be claimed since no long-range magnetic ordering has been reported in LCO \cite{Koehler1957, Thornton1986, Radaelli2002a, Belanger2016, Durand2015a, Durand2015}. 
Any materials disorder can strongly affect the long-range ordering, such as a coexisting Co$_3$O$_4$ phase reported to exist up to 5 wt \% in even high-quality LCO crystals \cite{Durand2015a, Durand2015}.
Surface FM in LCO is well known and can influence the magnetic ground state of bulk LCO depending on the materials preparation \cite{Senars-Rodrguez1995, Yan2004,  Durand2015}. Similarly, in LCO nanoparticles, an empirical model with an FM surface and an AFM core magnetism has been found to be the best explanation for the magnetic response \cite{Durand2015a}. 

We first introduce the magnetic phases of LCO that will be studied throughout this work. Detailed illustrations are given in Fig. \ref{fgr:structures}. Using the HS, IS, and LS spin states on Co$^{3+}$ ions, we investigate: G-type antiferromagnetic high-spin (HS-AFM), ferromagnetic high-spin (HS-FM), ferromagnetic intermediate-spin (IS-FM), ferromagnetic G-type mixed high-spin and low-spin (HS-FM/LS) and the low-spin (LS, nonmagnetic) states of LCO. The HS-FM, IS-FM, LS-FM, and HS-FM/LS phases have been studied previously using DFT \cite{Mukhopadhyay2013, Buckeridge2016, Korotin1996, Rondinelli2009, Ritzmann2014}. Therefore, these phases are included in our work as well for comparison. We have also included the HS-AFM phase, which is often not considered in other theoretical works. The magnetic phases we studied can be represented using the 10-atom unit cell of bulk LCO, which can form an acceptable starting point prior to studying supercells with more complex magnetic orderings.

In Fig. \ref{fgr:eos}, we present the DMC calculations with isotropic scaling in all three dimensions, to identify the DMC equilibrium volume and the DMC magnetic ground state of bulk LCO. 
In these calculations, we use the equilibrium LCO geometry from neutron diffraction experiments at 4 K (ICSD no. 201761)   \cite{Thornton1986} without applying geometry relaxation. In Fig. \ref{fgr:eos}, each fitted curve is obtained using the Murnaghan equation \cite{Murnaghan1944}. 
Fig. \ref{fgr:eos} shows that the HS-AFM state is predicted as the magnetic ground state. 
Given the recent experiments, which show stronger short-range AFM correlation up to 37 K \cite{Kaminsky2018b, Durand2013,Belanger2016, Itoh1995}, our results are worth further investigation.
At the experimental volume, HS-FM/LS is the second-most-stable phase according to DMC, followed by HS-FM. Compression yields a HS-FM/LS state more favorable compared to the HS-AFM state.
We find that the HS-FM and HS-AFM curves in Fig. \ref{fgr:eos} are almost parallel to each other within the investigated volume range. 
This can be explained by the isotropic scaling of the structure in Fig. \ref{fgr:eos}; any rotation on the Co-O octahedra and change in Co-O-Co angle is not allowed. 

In addition to the coexisting Co$_3$O$_4$ phase, the material may also have significant intrinsic defects in the powder form. In Section \ref{sec:excited}, we discuss that photoemission and optical conductivity measurements yield a much smaller band gap compared to the Kohn-Sham gaps from DFT functionals \cite{Chainani1992, Abbate1993, RazzaqueSarker2015, Tokura1998, Arima}. It is known that oxygen vacancies can change the charge state of Co atoms, modifying the magnetic structure of LCO. However, to our knowledge, a complete study of possible intrinsic defects in LCO has not yet been done. Additionally, it has been claimed that a spin-canted magnetic structure can also be energetically more favorable compared to the magnetic states we studied in this work \cite{Seo2012}.  

Finally, in this section, we investigate the DMC equilibrium volume of HS-AFM bulk LCO. 
Fig. \ref{fgr:eos} shows that HS-AFM has an equilibrium volume of 58.2(1) {\AA}$^3$ per formula unit (eV/f.u.), which is nearly 4$\%$ larger than the experimental volume of 55.8(3). 
DMC overestimates the equilibrium volume by nearly 4$\%$, hence 1.3$\%$ for the lattice parameter. The accuracy of DMC on the equilibrium volume is comparable to the DFT functionals (see Section \ref{sec:dft} and Table \ref{table:2}).
It was shown that the La pseudopotential we use overestimates the La$_2$O$_3$ equilibrium volume by nearly 3\%, although it produces excellent cohesive energies and bulk moduli with DMC \cite{Santana2016a}. 
Given that the La$^{3+}$ ionic radii is almost twice as large as Co$^{3+}$ \cite{Shannon1976}, the La$^{3+}$ ion can dominate the packing of the LCO crystal, and performance similar to La$_2$O$_3$ can be observed in equilibrium volume. 
We also calculated the standard formation enthalpy of LaCoO$_3$ using DMC and found a formation energy of 2.62(1) eV eV/f.u. compared to the experimental standard formation enthalpy of 2.55(1) eV/f.u. Details of this calculation are found in the Supplementary Information \cite{supp}. We must note that the DMC formation enthalpy calculation also yields the same HS-AFM phase. The HS-AFM phase is of lower energy compared to other magnetic phases; therefore, the other magnetic phases investigated yielded less accurate estimates of the formation energy up to 0.7 eV.
Further benchmarks with DFT functionals on the ground-state energies and volumes are given in Section \ref{sec:dft}.

\subsection{Epitaxial LCO} \label{sec:latmod}

In this section, we study the lattice modulation (i.e., superstructure formation) of the epitaxial LCO thin films on the strontium titanate (STO) substrate. We first examine how the simulation cells used in this section are constructed. LaCoO$_3$ is known to grow in a cube-on-cube manner on many substrates including STO, since STO also has a cubic lattice with $a_{STO}=3.905$ {\AA} \cite{Fuchs2007}. 
Therefore, we use a 2$\times$2$\times$2 (40-atom) pseudocubic cell of LCO using the starting ionic positions as mapped from the bulk LCO.
We define $a$ and $b$ as the in-plane lattice parameters, whereas $c$ is the out-of-plane lattice parameter. Similarly, $\overrightarrow{a}$, $\overrightarrow{b}$, and $\overrightarrow{c}$ are defined as the corresponding lattice directions. 
Using this starting cell, ionic relaxation is performed with PBEsol+$U$ = 4 eV, where $a=b=2\times3.905$ {\AA}, while the $c$ is varied systematically using fixed lattice angles (90$^{\circ}$). The out-of-plane axis lattice constant, $c$, is found to be 7.52 {\AA}, which agrees with the experiment (7.524 {\AA}) \cite{Mehta2015, Li2008} and PBE (7.54 {\AA}) \cite{Kwon2014a}. Therefore, $c$ is kept constant for all the remaining calculations performed in this section.

\begin{figure}[t]
	\includegraphics[width=0.5\textwidth]{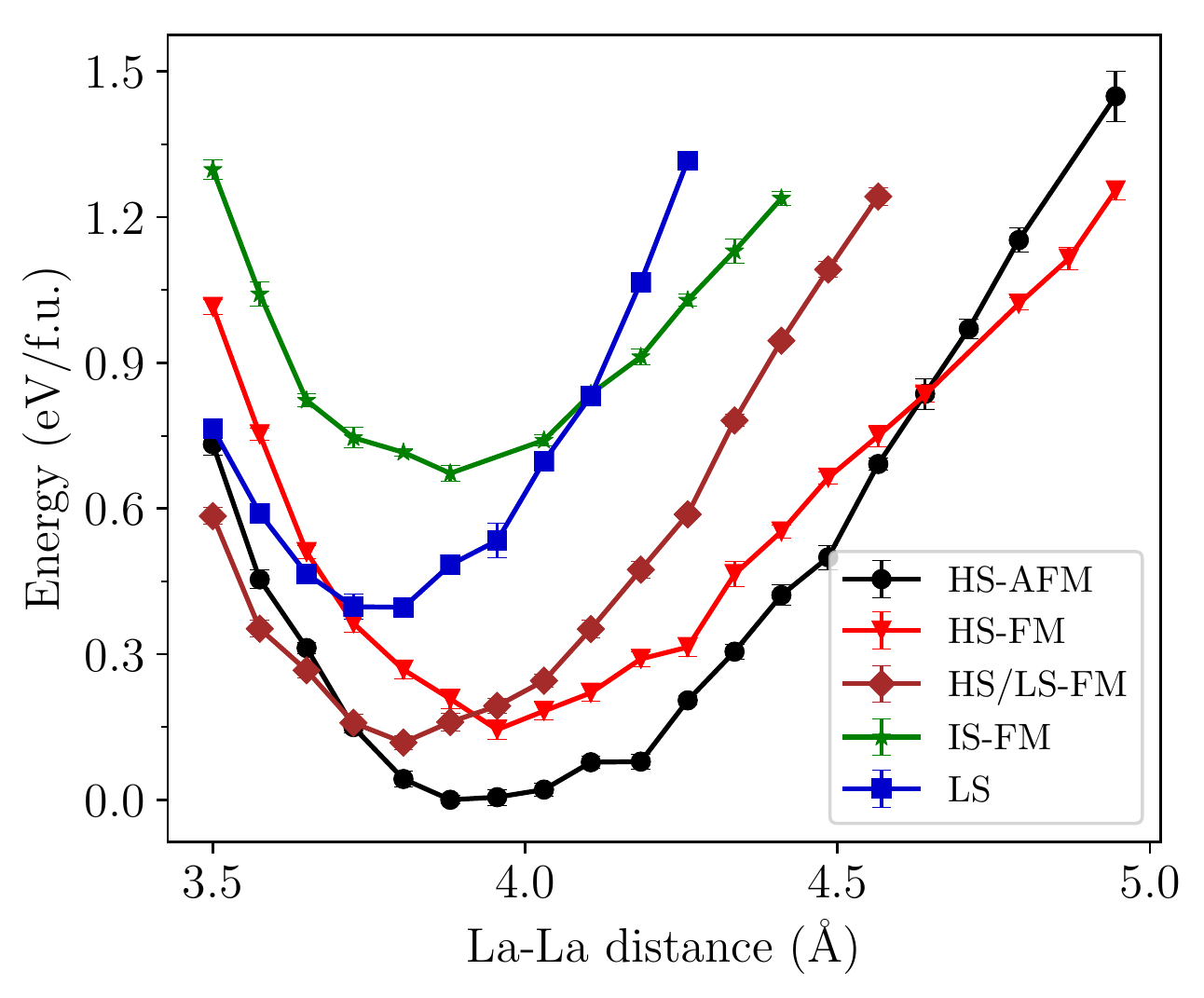}
	\caption{(Color online) DMC energy of epitaxial LCO/f.u. as a function of La-La distance, for the magnetic states given in Fig. \ref{fgr:structures} \label{fgr:uniaxial}. Energy on the y-axis is relative to the ground-state energy of HS-AFM phase in this figure.}
\end{figure}

\subsubsection{Uniform La-La Separation} \label{sec:latmod1}
We initially studied the system with uniform La-La separation along the $\overrightarrow{a}$ direction, by varying the size of $a$, while keeping the remaining lattice parameters constant. Meanwhile, all the ionic degrees of freedom are optimized, except for the La atoms in the $\overrightarrow{a}$ direction.  The geometry optimization was performed separately for each point in Fig. \ref{fgr:uniaxial}. Our results are presented in Fig. \ref{fgr:uniaxial}. We compare the results in Fig. \ref{fgr:uniaxial} to Fig. \ref{fgr:eos} and find that the results are qualitatively similar near each equilibrium. 
However, the results in Fig. \ref{fgr:uniaxial} are more scattered compared to Fig. \ref{fgr:eos}, due to the systematic contributions from the DFT relaxations, as previously observed in \cite{Rondinelli2009}.
The energy difference between the minimum energy LS and HS-AFM structures is identical in Fig. \ref{fgr:uniaxial} and Fig. \ref{fgr:eos} with 0.40(2) and 0.39(1) eV. 
However, the difference between the HS-AFM and the HS-FM minima reduces from 0.22(1) in Fig. \ref{fgr:eos} to 0.14(1) eV in Fig. \ref{fgr:uniaxial}.  
A similar reduction is also observed between the HS-AFM and the HS-FM/LS states, from 0.20(2) to 0.11(1) eV.
Because the stability between HS-AFM and LS states remains unchanged, we can conclude that HS-FM and HS-FM/LS states become more stabilized with geometry optimization in Fig. \ref{fgr:uniaxial}.
We find that the ionic relaxations on the Co and O atoms under uniaxial strain lead to distorted octahedra, compared to the perfect octahedra of the experimental structure we studied in Fig. \ref{fgr:eos}. However, qualitative energetic ordering around the equilibrium in each case is unchanged between the two studies.

Another important result from Fig. \ref{fgr:uniaxial} is the crossing between the HS-FM and HS-AFM energy curves at 4.6 {\AA}. A crossing between HS-FM and HS-AFM states in Fig. \ref{fgr:uniaxial} indicates that exchange coupling the constant changes its sign as a function of the La-La separation.  Goodenough-Kanamori-Anderson (GKA) rules   \cite{Goodenough1958a, Kanamori1960, Anderson1950} state that $\sim$180{$^{\circ}$} superexchange of two magnetic ions with partially filled $d$ orbitals is strongly antiferromagnetic. However, a FM state can be stabilized over the an AFM state if orbital ordering exists in the FM state \cite{Kugel1982}. This point will be discussed further in Section \ref{sec:oo}. 
The calculations in Fig. \ref{fgr:uniaxial} are useful to show that (1) a crossing between HS-FM and HS-AFM states is possible as a function of strain and (2) the qualitative energetic ordering is identical both in Fig. \ref{fgr:eos} and \ref{fgr:uniaxial}. 
In Fig. \ref{fgr:uniaxial}, we have only studied a single magnetic phase throughout the bulk material. 
However, bright and dark stripes shown in the STEM images of epitaxial LCO samples \cite{Biskup2014, Mehta2015,Gazquez2011a, Kwon2014a, Choi2012, Qiao2015} clearly demonstrate a structural modification with two lateral domains. 
These domains can choose different magnetic ground states given the large variation in their La-La separations. 
Therefore, in Section \ref{sec:latmod2}, we will study the structural modulation of epitaxial LCO along with the different magnetic phases applied on each lateral domain.

\begin{figure}[t]
	\includegraphics[width=0.5\textwidth]{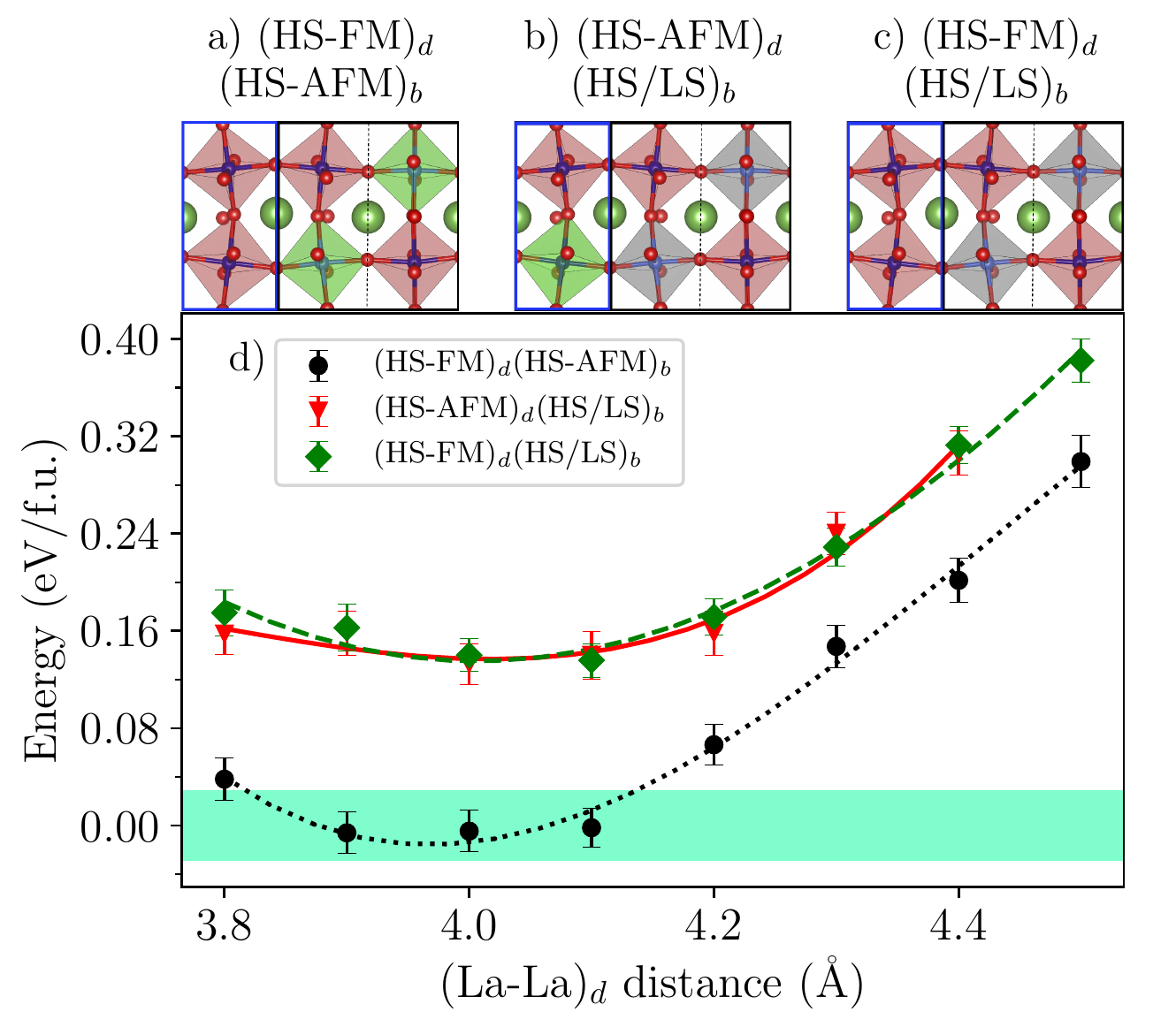}
	\caption{(Color online) Lattice modulation in epitaxial LCO. (a)--(c) Magnetic configurations of the superstructures. Superstructures are formed in a single larger striped and two smaller striped regions, which are identified with their in-plane La-La distances. Larger stripes are associated with dark, $d$, regions, and smaller stripes are associated with bright, $b$, regions as seen in their STEM analysis. The coloring of the atoms is identical to Fig. \ref{fgr:structures}. Additionally, La atoms are shown in green. (d) DMC energies of the magnetic configurations in (a)--(c) as a function of La-La separation of the dark striped regions. Energies on the y-axis are relative to the HS-AFM phase in Fig. \ref{fgr:uniaxial} with identical amount of strain. \label{fgr:lm}}
\end{figure}

\subsubsection{Modulated La-La Separation} \label{sec:latmod2}
In Fig. \ref{fgr:lm}, we study the lattice modulation under the epitaxial conditions with two coexisting magnetic phases. The definitions for the lattice parameters and directions here are identical to Section \ref{sec:latmod}. 
Here, the lattice parameters are fixed as $a=3\times{a_{STO}}$, $b=2\times{a_{STO}}$, and $c=7.52$ {\AA}. However, the La-La distances along the $\overrightarrow{a}$ direction are modulated to simulate the superstructure with two bright stripes and one dark stripe as seen in the STEM images \cite{Biskup2014, Mehta2015,Gazquez2011a, Kwon2014a, Choi2012, Qiao2015}. 
To simulate this structure, we use a relation such as (La-La)$_d$ + 2 $\times$ (La-La)$_b = a$. Here, (La-La)$_d$ is the La-La separation in the dark stripes, whereas (La-La)$_b$ is the La-La separation in the bright stripes, both along the $\overrightarrow{a}$ direction. 
In Fig. \ref{fgr:lm}, (La-La)$_d$ is varied between 3.8 to 4.5 {\AA}. 
La ionic degrees of freedom along the $\overrightarrow{a}$ direction are kept fixed throughout the geometry optimization to maintain the lateral modulation. Geometry optimizations are performed separately for each point in Fig. \ref{fgr:lm}. 

We study three different spin configurations in the lattice-modulated superstructure defined above. 
The spin configurations we studied are shown in Fig. \ref{fgr:lm}(a)--(c). 
Here, the up-spin Co$^{3+}$ octahedra are shown with red octahedra, down-spin Co$^{3+}$ octahedra are shown in green, and the nonmagnetic (LS) Co$^{3+}$ octhedra are shown in gray. 
For example, (HS-FM)$_d$(HS-AFM)$_b$ means HS-FM ordering is in the dark (larger, single-striped) region, whereas HS-AFM ordering is studied in the bright (smaller, double-striped) region as seen in the STEM images. 
In Fig. \ref{fgr:lm}, energies on the y-axis are relative to the HS-AFM phase with uniform La-La separation (as in Fig. \ref{fgr:uniaxial}) for every point.
All magnetic states studied in Fig. \ref{fgr:lm}, including (HS-FM)$_d$(HS-AFM)$_b$, have an optimal (La-La)$_d$ distance of $\sim$4.0 {\AA}, meaning (La-La)$_b$ is $\sim$3.85 {\AA}, which is slightly larger than the lattice parameter of the STO unit cell (3.905 {\AA}).
We list the results obtained from Fig. \ref{fgr:lm}(d) below for a better discussion. 
\begin{enumerate}[label={\roman*)}, noitemsep]
    \item  It costs almost no energy to induce a lattice modulation with (HS-FM)$_d$(HS-AFM)$_b$ magnetic structure under epitaxial strain. STEM images from the LCO on STO samples \cite{Kwon2014a, Biskup2014, Mehta2015} show that the immediate LCO on STO boundary has very small lattice modulation. Although most studies focus on the emergence of ferromagnetism within the LCO layer (which has much larger lattice modulation), we show that a FM layer along the LCO/STO boundary can also be energetically favorable. 
    \item To stabilize experimentally observed (La-La)$_d$ (4.5 {\AA}), an additional 0.3 eV/f.u. is required. 
    \item Substitutional or vacancy defect formations may be needed to drive lattice separations, (La-La)$_d$, of the magnitude observed in experiments. Oxygen vacancy formation has been suggested as a likely defect in the dark-striped regions, but this would mean a change from Co$^{3+}$ to Co$^{2+}$ charge state as well. So far, this has not been observed in the Co electron energy loss spectra  \cite{Kwon2014a, Biskup2014, Mehta2015}. 
    \item As the (La-La)$_d$ distance is increased, energies of the distinct configurations start approaching each other with less than 2 $\sigma$ uncertainty, with $\sigma <$ 0.1 eV. Meaning that the magnetic configurations not studied in Fig. \ref{fgr:lm} could also yield very similar energies at (La-La)$_d$ separation of 4.5 {\AA}. 
\end{enumerate}

\begin{table}
	\caption{\label{table:2} Excited-state properties of LS and HS-AFM LCO. Indirect and direct band gaps from DFT calculations correspond to the generalized Kohn-Sham eigenvalue differences from band-structure calculations. DMC$_{QP}$ is the quasiparticle gap calculated using DMC. DFT values reported from the literature are obtained using the density of states calculations. The asterisk ($*$) indicates that the calculations were performed using VASP-PAW pseudopotentials, whereas the dagger ($\dagger$) indicates that the QE-RRKJ pseudopotentials were used.}
	\begin{ruledtabular}
		\begin{tabular}{l c}
\textit{HS-AFM} &	$E_g (eV)$ 	\\
\hline
LDA(+$U$=6 eV)	&	1.94$^\dagger$	\\
DMC	(X$\rightarrow$X) & 3.77 $\pm$ 0.12	\\
DMC$_{QP}$	(X$\rightarrow$X) & 3.87 $\pm$ 0.22  \\
\textit{LS}  &	\\
\hline
LDA(+$U$=4 eV)	&	0.72$^*$, 0.87$^*$ \cite{Ritzmann2014}, 0.95$^\dagger$	\\
LDA(+$U$=6 eV)	&	1.15$^*$, 1.45$^\dagger$ \\
LDA(+$U$=7 eV)	&	1.23$^*$, 1.6$^\dagger$, 1.72 \cite{Hsu2009}	\\
LDA(+$U$=7.8 eV, $+J$=0.9 eV) & 1.23$^*$, 2.06   \cite{Korotin1996} \\
PBE(+$U$=4 eV)	&	1.12$^*$, 1.12$^*$\cite{Ritzmann2014}, 	\\
PBE(+$U$=5.4 eV)	&	1.25$^*$, 1.58$^\dagger$, 1.5 \cite{Knizek}	\\
PBE(+$U$=6 eV)	&	1.34$^*$, 1.7$^\dagger$	\\
B3LYP & 2.2   \cite{Mukhopadhyay2013} \\
HSE & 2.38$^*$, 2.44 \cite{He2012a}, 2.54$^\dagger$ \\
PBE0 & 2.42$^*$, 2.4$^*$ \cite{Gryaznov}, 3.29$^\dagger$, 3.2\cite{Gryaznov} \\
DMC	($\Gamma\rightarrow{\Gamma}$) & 3.65  $\pm$ 0.06	\\
DMC$_{QP}$	($\Gamma\rightarrow{\Gamma}$) & 3.7 $\pm$ 0.1  \\
\textit{Experimental} & \\
\hline
Optical conductivity & 0.1-1.1   \cite{RazzaqueSarker2015, Tokura1998, Arima} \\
X-ray photoemission spectroscopy& 0.6-0.9   \cite{Chainani1992, Abbate1993} \\
Photoluminescence and UV/Vis & 3.44-3.50 \cite{Jayapandi2018a, Jayapandi2018b}\\
		\end{tabular}
	\end{ruledtabular}
\end{table}

\subsection{Electronic structure, quasiparticle and optical gaps}\label{sec:excited}

In Table \ref{table:2}, we show the experimental, DFT Kohn-Sham, and DMC optical/quasiparticle gaps for the HS-AFM and LS states of LCO. 
However, before presenting the DMC band gaps, we first benchmark the DFT band gaps of LS LCO using our calculations and the results collected from the literature. Because QE-RRKJ orbitals are used to perform ground-state and excited-state DMC calculations, it is important to benchmark their performance with respect to the other codes and the pseudopotentials to test any systematic difference. The results with an asterisk ($^*$) in Table \ref{table:2} are VASP-PAW calculations, whereas all the calculations with ($^\dagger$) are using QE-RRKJ. 
All electron band gap calculations using WIEN2k-FLAPW \cite{WIEN2k, Hsu2009, Knizek}, CRYSTAL-LCAO \cite{Dovesi2018}, and LMTO \cite{Korotin1996} are taken from the literature. 
Unless a reference is given next to a value, all the calculations in Table \ref{table:2} are performed by us.
A general trend we identified in these DFT calculations is that harder pseudopotentials \cite{Krogel2016} and all-electron calculations yield larger band gaps compared to softer core pseudopotentials in VASP-PAW \cite{Kresse1999}. In LDA+($U=$ 7 eV) calculations, VASP-PAW yields a gap of 1.23, whereas a gap with 1.72 eV \cite{Hsu2009} is found using WIEN2k-FLAPW. Similarly, for LDA$+U = $7.8 eV and $+J = $0.9 eV calculations (using Liechtenstein's rotationally invariant method \cite{Liechtenstein1995}), VASP-PAW underestimates the band gap by nearly 0.8 eV compared to LMTO \cite{Korotin1996}. We found that the VASP LDA$+U+J$ band gap is identical to the LDA$+U=$ 7 eV band gap (1.23 eV) using Dudarev's simplified scheme ($U_{eff}=U-J$), meaning that the $+U$ implementation has only minimal effect on the band gaps. 
Interestingly, a large discrepancy was reported by Gryaznov et. al. \cite{Gryaznov} between the PBE0 VASP-PAW and CRYSTAL-LCAO band gaps (2.4 eV vs. 3.2 eV). Our VASP-PAW DFT+$U = $4 eV and hybrid-DFT calculations are in very good agreement with the literature; therefore, we validate the VASP-PAW results from Gryaznov et. al. \cite{Gryaznov}. 
On the other hand, the benchmark on the QE-RRKJ band gaps indicates that QE-RRKJ band gaps are closer to the WIEN2k-FLAPW gaps with respect to the VASP-PAW gaps. 
The QE-RRKJ PBE0 band gap is in very good agreement with the CRYSTAL-LCAO, again indicating an estimate that is 0.8 eV larger than the VASP-PAW PBE0 band gap. 

The hybrid-DFT band gap values are rather large compared to some of the experiments claiming 0.6--0.9 eV  from photoemission measurements \cite{Chainani1992, Abbate1993}, and at 0.1--1.1 eV  from optical conductivity measurements \cite{RazzaqueSarker2015, Tokura1998, Arima}.  However, the recent photoluminescence and UV-Vis spectroscopy measurements yield a gap of nearly 3.5 eV which agrees very well with the CRYSTAL-LCAO and QE-RRKJ PBE0 calculations. There can be multiple reasons to explain this discrepancy: (1) The interpretation of the experimental spectra has been challenging. While a value of 0.6 eV is reported by Chainani et. al. \cite{Chainani1992}, the same data are interpreted as 2--3 eV by Saitoh et. al. \cite{Saitoh1997}.
(2) It is also possible that some of the measured excitations may correspond to internal $d$-$d$ transitions.
We have reported a similar discrepancy in the band gap of CoO, where the band gaps were found to be between 2.5 and 6 eV using different experimental techniques \cite{Saritas2018b}. 
However, in CoO, ellipsometry studies have indicated that the band gap is observed at 5.43 eV, while the lower energy excitations (around 2--3 eV) may correspond to the internal $d$-$d$ transitions \cite{Dong2007}. 
(3) The flexible nature of the magnetic state of LCO may also play a role in the discrepancy between some of the theoretical and experimental results.
Fig. \ref{fgr:eos} shows that different magnetic states of Co$^{3+}$ (e.g., HS or IS) can be accessible within 0.7 eV per Co of the ground state. 
This may suggest that changing the spin of a single Co$^{3+}$ ion can be achieved within an excitation of similar magnitude. 
Therefore, we believe that additional experimental studies may be needed to understand the optical transitions in LCO. 

\begin{figure}[t]
	\includegraphics[width=0.5\textwidth]{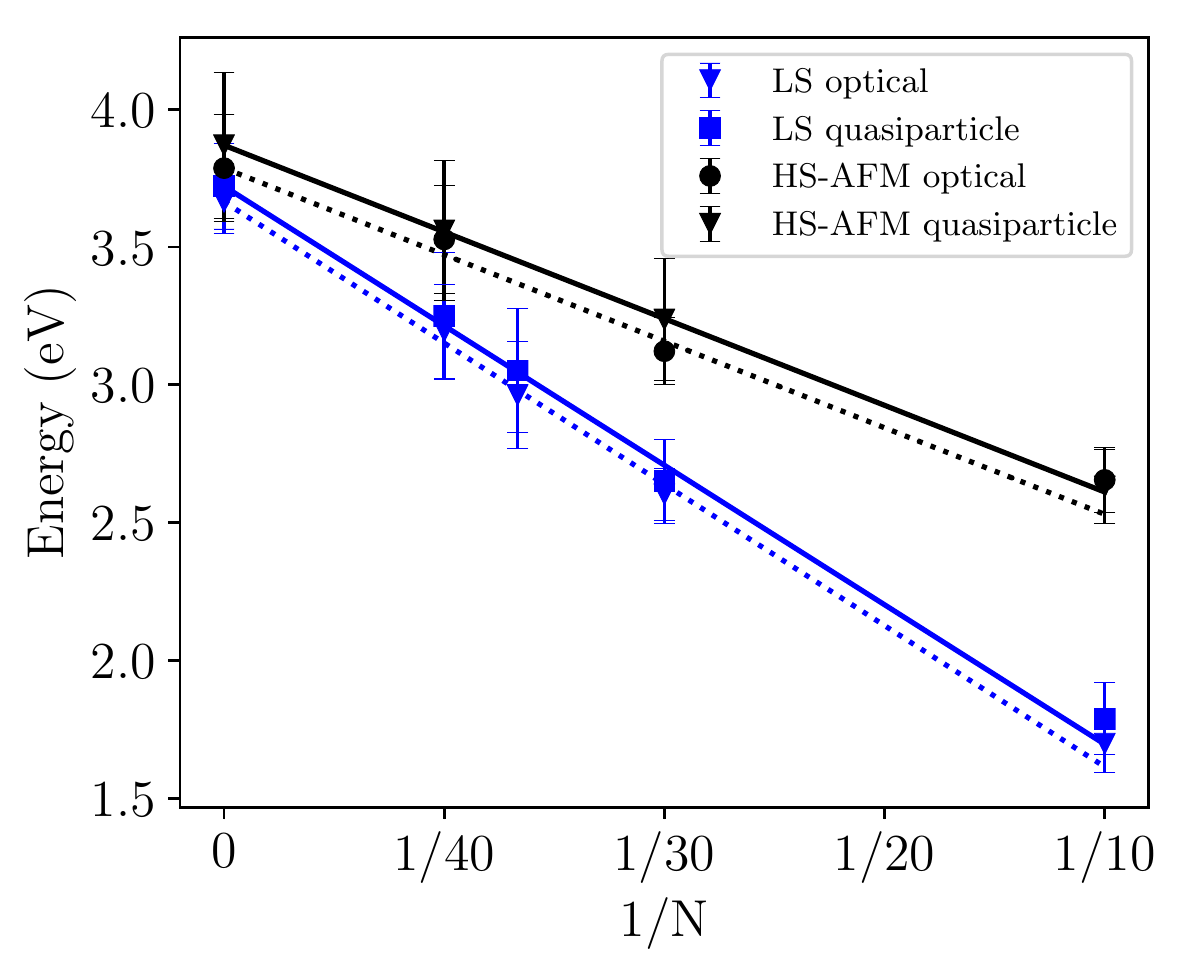}
	\caption{(Color online) Optical and quasiparticle gaps of LS and AFM LCO using DMC. For LS and AFM states, $\Gamma\rightarrow$F and L$\rightarrow$F transitions are investigated, respectively.  The x-axis is the inverse number of atoms in the simulation cell, while the y-axis is the gap energies in eV. LS and AFM energies are shown in blue and black, respectively. Optical gaps are connected with dashed lines, while quasiparticle gaps are connected with solid lines. \label{fgr:dmc_excited}}
\end{figure}

Excited-state DMC calculations require identifying the conduction band and valence band wavevectors of the excitation. 
Therefore, we obtained band structures using LDA$+U$=6 eV to identify VBM and CBM. 
The band structures of both materials are given in the Supplementary Information \cite{supp}.
We find that both HS-AFM and LS LCO are indirect band gap materials, where indirect and direct band gap energies differ less than 0.1 eV in the LDA+$U$ calculations. HS-AFM LCO has a well defined valence band maximum at $X$ with a bandwidth of 1 eV at the valence band. However, HS-AFM LCO conduction band and the LS LCO conduction and valence bands are found to be rather flat with bandwidths smaller than 0.2 eV.
Therefore, we study the direct band gaps in the DMC calculations, since it provides more flexible choice for the supercell tiling matrices to eliminate the finite size errors. We study HS-AFM direct band gap at $X$, while the LS-LCO is studied at $\Gamma$. 
The reported DMC band gap values are obtained from the finite-size extrapolation in Fig. \ref{fgr:dmc_excited}. 
We find DMC band gaps of roughly $\sim$3.7(2) eV for both LS and HS-AFM LCO. This agrees very well with the photoluminescence and UV/Vis \cite{Jayapandi2018a, Jayapandi2018b} experiments and also with PBE0, where a band gap of 3.2--3.3 eV is obtained using QE-RRKJ and CRYSTAL-LCAO. DMC band gaps are larger than the hybrid-DFT band gaps as observed previously  \cite{Saritas2018a, Saritas2018b, Kent1999}. We used a simple single-determinant scheme to obtain the band gaps from DMC; therefore, our results should be treated as an upper bound due to the fixed node bias, which may not fully cancel between the ground and excited states. Multideterminant wavefunctions can be used to optimize the excited-state nodal surface and control fixed node bias, although studies for extended systems are very limited \cite{Scemama2018a, Blunt2019, Blunt2017}, because of the significant computational resources that would be required using DMC. 
A very recent work on VMC, however, shows that nodal surface errors can be minimized using orbital rotations on the single-determinant wavefunction \cite{Zhao2019}. DMC quasiparticle and optical gaps in Fig. \ref{fgr:dmc_excited} are identical given the statistical uncertainties, suggesting very small exciton binding energies.

\subsection{Orbital ordering in LCO}\label{sec:oo}
Superexchange interactions usually lead to strong AFM when the transition metal $d$-orbitals containing single electrons overlap over the intermediate anions (ligands) near linear angles  \cite{Goodenough1958a, Kanamori1960, Anderson1950}. However, in broken symmetry states, orbital ordering may emerge, leading to an alternating pattern of localized occupied orbitals   \cite{Pavarini2016}. 
If orbital ordering occurs, the filled orbital at one site can overlap with a vacant orbital in the adjacent site and lead to relatively weaker FM interactions   \cite{Kugel1982}. 
Orbital ordering in LCO was initially proposed for the formation of the IS-FM  state   \cite{Korotin1996}. 
Co$^{3+}$ ions have the $t_{2g}^5e_g^1$ configuration in the IS-FM state where the ordering is observed on the majority-spin $e_g$ orbitals. 
In HS-FM LCO, however, the majority-spin orbitals on Co$^{3+}$ are completely filled. 
Therefore, the orbital ordering can only form over the minority spin $t_{2g}$ orbitals as the hopping between the filled parallel spin electrons is forbidden by the Pauli principle. The superexchange mechanism among the minority spin electrons of the Co$^{3+}$ ions in the HS-FM phase LCO is analogous to the superexchange of $d^1$ ions such as Ti$^{3+}$ in LaTiO$_3$ and YTiO$_3$. 

\begin{figure}[t]
	\includegraphics[width=0.5\textwidth]{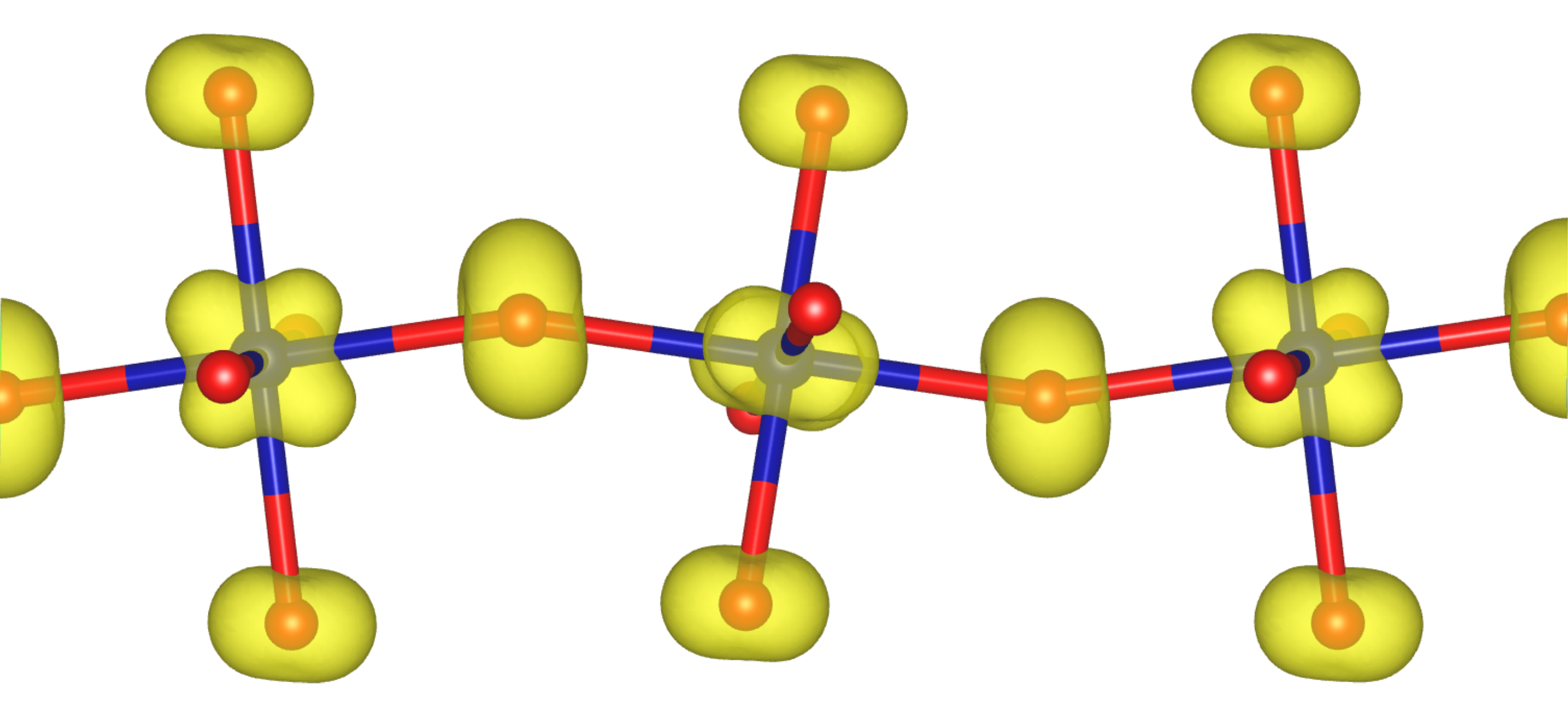}
	\caption{(Color online) Orbital ordering in the high-spin FM state for the occupied minority $t_{2g}$ orbitals, hybridized with O-$p$, along the  [100] direction of the pseudocubic cell. La atoms and the periodicity in $y$ and $z$ directions are omitted for clarity. \label{fgr:oo}}
\end{figure}

In Fig. \ref{fgr:oo}, we plot the band decomposed charge density at the VBM of the minority spin electrons for HS-FM LCO to demonstrate the orbital ordering. The ordering alternates between the d$_{xy}$ and d$_{xz}$ orbitals. 
As expected, a similar orbital ordering has not been observed for the HS-AFM state. 
In the case of orbital ordering, GKA rules do not apply, and the hopping from a filled to a degenerate empty orbital favors the FM interaction over the AFM interaction. 
The presence of orbital ordering is also supported by the difference in octahedra distortion indices  \cite{Baur1974} found on the Co-O octahedra. 
We find that the distortion index monotonously increases in both HS-FM and HS-AFM states as a function of increased uniaxial strain, while the distortion index is less than 2\% near the equilibrium of each curve.

\subsection{Comparisons with DFT methods} \label{sec:dft}

Several theoretical approaches have been used to study the role of various external stimuli and the environment of the Co atom on the spin state of bulk and epitaxially strained LCO. 
Density functional mean field   \cite{Krapek2012, Izquierdo2014} and quantum chemistry calculations   \cite{Wang2015, Siurakshina2010, Eder2010, Takahashi1998, Saitoh1997} have been extensively used for this end   \cite{Hsu2009,Kushima2010, Han2011, Mukhopadhyay2013, Ritzmann2014}. 
DFT methods with Hubbard-$U$   \cite{Anisimov1991a, Dudarev1998} corrections and exchange mixing   \cite{Heyd2003, Becke1993b} are often highly tuned to reproduce bulk spin ground states, band gaps, and geometric properties reported experimentally   \cite{Buckeridge2016, Rondinelli2009}. The accuracy of these methods depends critically on the corrections applied. Experimental properties might also depend strongly on the presence of defects and strain.
Therefore, the use of a more accurate method, such as DMC, and benchmarking is rather critical for consistent results. 

In Table \ref{table:1}, we present our benchmark on the ground-state energetic, magnetic, and structural properties of LCO.
Here, the geometry relaxations are performed for the DFT calculations, whereas DMC results are transferred from Fig. \ref{fgr:eos}. 
Table \ref{table:1} shows that, with increased $U$ values, the HS-AFM state is stabilized over the LS state for all DFT functionals. 
For LDA, the crossing between LS and HS-AFM states occurs at a $U$ value of 4--6 eV, whereas for PBE this crossing occurs with a smaller value of $U$ between 0--2 eV. 
However, with SCAN, the $+U$ correction is not needed to stabilize the HS-AFM state over the LS state. 
We find that hybrid-DFT functionals, HSE and B3LYP, also predict the HS-AFM state to be more stable than the LS state.
Interestingly, Table \ref{table:1} shows that it is not possible to stabilize the LS state over the HS-AFM state with the SCAN+$U$ approach. 
This contrasts with  LDA and PBE, where tuning the +$U$ parameter to smaller values allows for studying the LS state as the ground state of LCO \cite{Korotin1996}. 
Various examples in the literature suggest a reduced self-interaction error and an improvement in performance with SCAN over GGA \cite{Perdew2017, Peng2017, Kitchaev2016, SaiGautam2018a}. 
This is also observed in our calculations such that significantly lower $U$ values are required with SCAN to reproduce equivalent results with GGA+$U$, indicating a possible improved description of the exchange interactions is provided with SCAN. 
We previously observed that SCAN provided an improvement over PBE in predicting the energy differences between CoO polymorphs \cite{Saritas2018b}.

\begin{table}
	\caption{\label{table:1} Stability of the HS-AFM state with respect to the LS state, $\Delta{E}=E_{HS-AFM}-E_{LS}$, ground-state magnetic structure and ground-state volume percentage error from DFT functionals. $\Delta{V}=V-V_0$, where $V$ is the equilibrium volume from the used method and $V_0$ is the experimental volume. Energies are given eV per formula unit, and $U$ values used are in units of eV. For $\Delta{E}$, no numerical value indicates that HS-AFM state is not stable. }
	\begin{ruledtabular}
		\begin{tabular}{l c c c}
	&	$\Delta{E}$	(eV) &	Ground state	&	$\Delta{V}/{V_0}*10^2$ 	\\
	\hline
LDA	&	-	&	LS	&	-7.16	\\
LDA(+$U$=2 eV)	&	0.52	&	LS	&	-7.51	\\
LDA(+$U$=4 eV)	&	0.21	&	LS	&	-7.70	\\
LDA(+$U$=6 eV)	&	-0.19	&	HS-AFM	&	-2.54	\\
LDA(+$U$=8 eV)	&	-0.65	&	HS-AFM	&	-4.02	\\
PW91	&	-	&	LS	&	-0.32	\\
PBEsol	&	-	&	LS	&	-4.52	\\
PBEsol(+$U$=4 eV)	&	-0.19	&	HS-AFM	&	1.70	\\
PBE	&	-	&	LS	&	-0.21	\\
PBE(+$U$=2 eV)	&	-0.04	&	HS-FM/LS	&	2.56	\\
PBE(+$U$=4 eV)	&	-0.38	&	HS-AFM	&	5.78	\\
PBE(+$U$=6 eV)	&	-0.77	&	HS-AFM	&	5.40	\\
PBE(+$U$=8 eV)	&	-1.20	&	HS-AFM	&	4.50	\\
SCAN	&	-0.77	&	HS-AFM	&	3.50	\\
SCAN(+$U$=2 eV)	&	-1.43	&	HS-AFM	&	3.55	\\
SCAN(+$U$=4 eV)	&	-1.95	&	HS-AFM	&	2.77	\\
SCAN(+$U$=6 eV)	&	-2.45	&	HS-AFM	&	2.06	\\
SCAN(+$U$=8 eV)	&	-2.99	&	HS-AFM	&	1.08	\\
B3LYP	&	-0.09	&	HS-AFM	&	7.80	\\
HSE	&	-0.19	 &	HS-AFM	&	-4.30	\\
DMC	&	-0.40(2)	 &	HS-AFM	&	4.30(5)	\\
		\end{tabular}
	\end{ruledtabular}
\end{table}

Percent errors for the equilibrium volumes are also presented for each method in Table \ref{table:2}. As expected, LDA largely underestimates the equilibrium volume, while PBE underestimates with a much smaller percentage. PW91, PBEsol+$U$ = 4 eV, and SCAN+$U$ = 8 eV have are the most accurate in reproducing equilibrium volume. The performance of the PBEsol+$U$ = 4 eV functional has been reported elsewhere for bulk LCO  \cite{Belanger2016}. As explained previously, the PBEsol+$U$ = 4 eV functional is used throughout this work to optimize the systems' structural parameters due to the functional's superior performance. Considering all the factors investigated in Table \ref{table:1}, we find that PBE+$U$ = 4 eV gives the best compromise in all properties compared to DMC. 
$U$ values of 3--4 eV were reported to be typical for PBE and the PAW pseudopotentials we use in binary Co-oxides, in terms of giving a reasonable compromise across different properties \cite{Seo2012, Wang2006a, Jain2011}. 

\section{Conclusions}\label{sec:conc}
DMC has consistently produced accurate structural and energetic properties for challenging materials problems, such as improving DFT approximations on transition-metal oxides. Because the accuracy of DMC is established in materials with minimal controversy between the experiments, we believe our results are significant enough to explain the experimental and theoretical controversies observed for various physical properties of LCO.

We studied bulk and epitaxial LCO subject to uniaxial strain and lattice modulation. We first found that bulk LCO has an AFM ground state that is of lower energy compared to the nonmagnetic state. This is in contrast to long-standing experiments; our experiments support this idea. However, experimental characterization of the magnetic ground state of LCO is a challenging problem that has been actively researched since the 1950s \cite{Koehler1957}. In our calculations, we also find that a magnetic state with net FM can be stabilized through uniaxial strain or compression. However, our study shows that the periodic structural deformation observed experimentally is not stabilized by the FM state. We showed that additional external factors, with an energy comparable to 0.3 eV/f.u., should play a role in generating lattice modulation of epitaxial LCO with experimentally observed La-La separation.

We calculated the optical and quasiparticle gaps of LCO and found that DFT and DMC predictions are significantly larger than the experimental values. We explained this discrepancy and showed that the spin transitions below 0.7 eV may explain the low excitation energies observed in the experiments. In addition, we pointed out that possible defects in the structure should also be considered. A subsequent study involving defects is under way.

In summary, we calculated the electronic, structural, and magnetic properties of bulk and epitaxial LCO using a range of density functionals and the DMC method to study the origin of FM in LCO thin films. We found the ground state of bulk LCO to be magnetic, while the G-type HS-AFM structure was the lowest energy among the structures considered. We discussed the significance of this result in light of recent experimental results and showed that, under epitaxial strain, a FM phase can appear with little extra energy and small lattice distortion. We pointed out that defects can provide sufficient energy to yield experimentally observed La-La separations in epitaxial LCO that is nearly 4.5 {\AA}. 
We found that DFT and DMC electronic gaps are significantly larger than experimental energy gaps, suggesting possible intrinsic defects.

\begin{acknowledgments}
We are grateful for helpful discussions with Dr. Jaime Fernandez-Baca and Dr. Feng Ye on analyzing our neutron diffraction experiments. The work was supported by the Materials Sciences and Engineering Division of the Office of Basic Energy Sciences, US Department of Energy (DOE). Computational resources were provided by the Oak Ridge Leadership Computing Facility at the Oak Ridge National Laboratory, supported by the Office of Science of the DOE under Contract no. DE-AC05-00OR22725. 
\\
\end{acknowledgments}

\appendix*

\section{DMC wavefunction optimization and convergence tests}


\begin{figure}[h] \label{fgr:ktest}
	\includegraphics[width=0.5\textwidth]{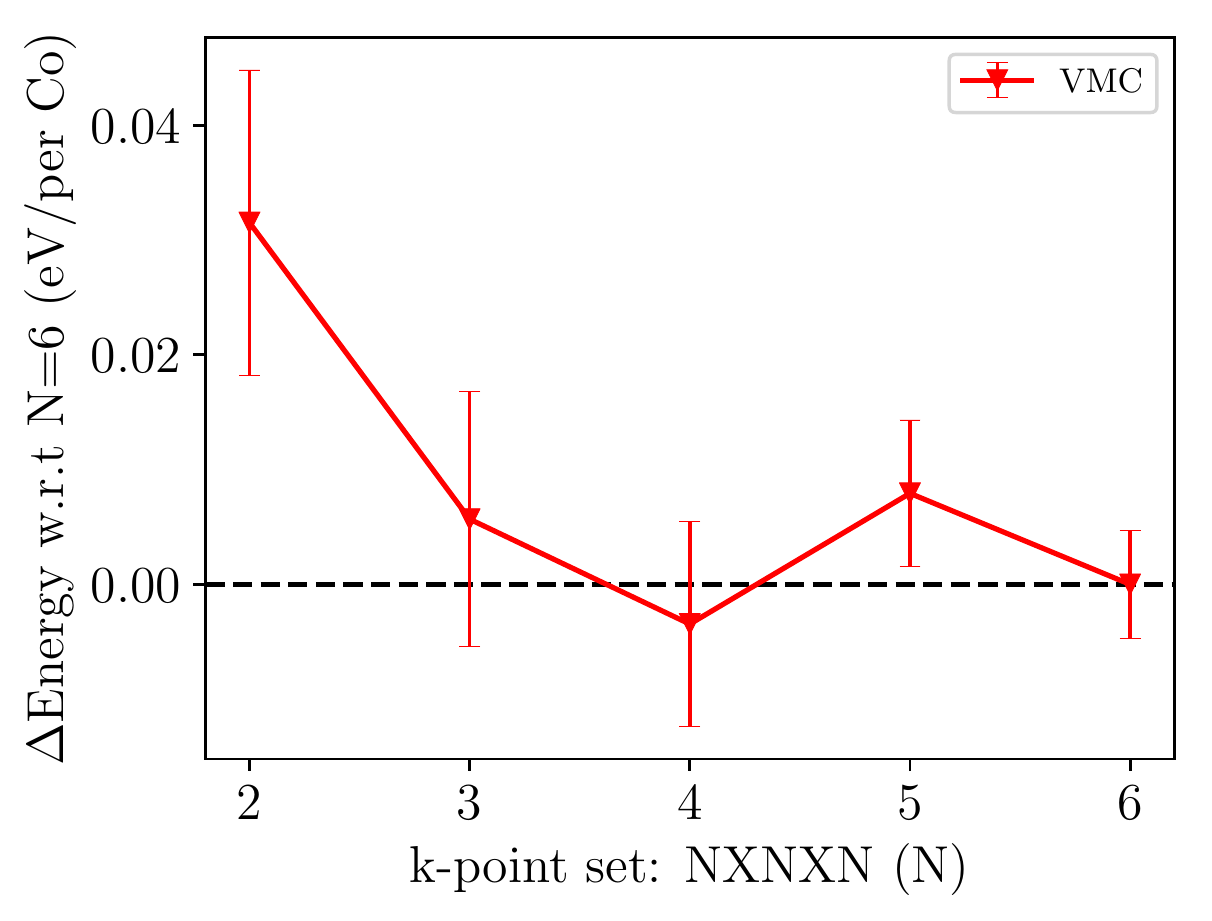}
	\caption{One-body finite size error as a function of k-point grid set size on 20-atom HS-AFM bulk LaCoO$_3$ using VMC. }
\end{figure}

\begin{figure}[h] \label{fgr:dfttest}
	\includegraphics[width=0.5\textwidth]{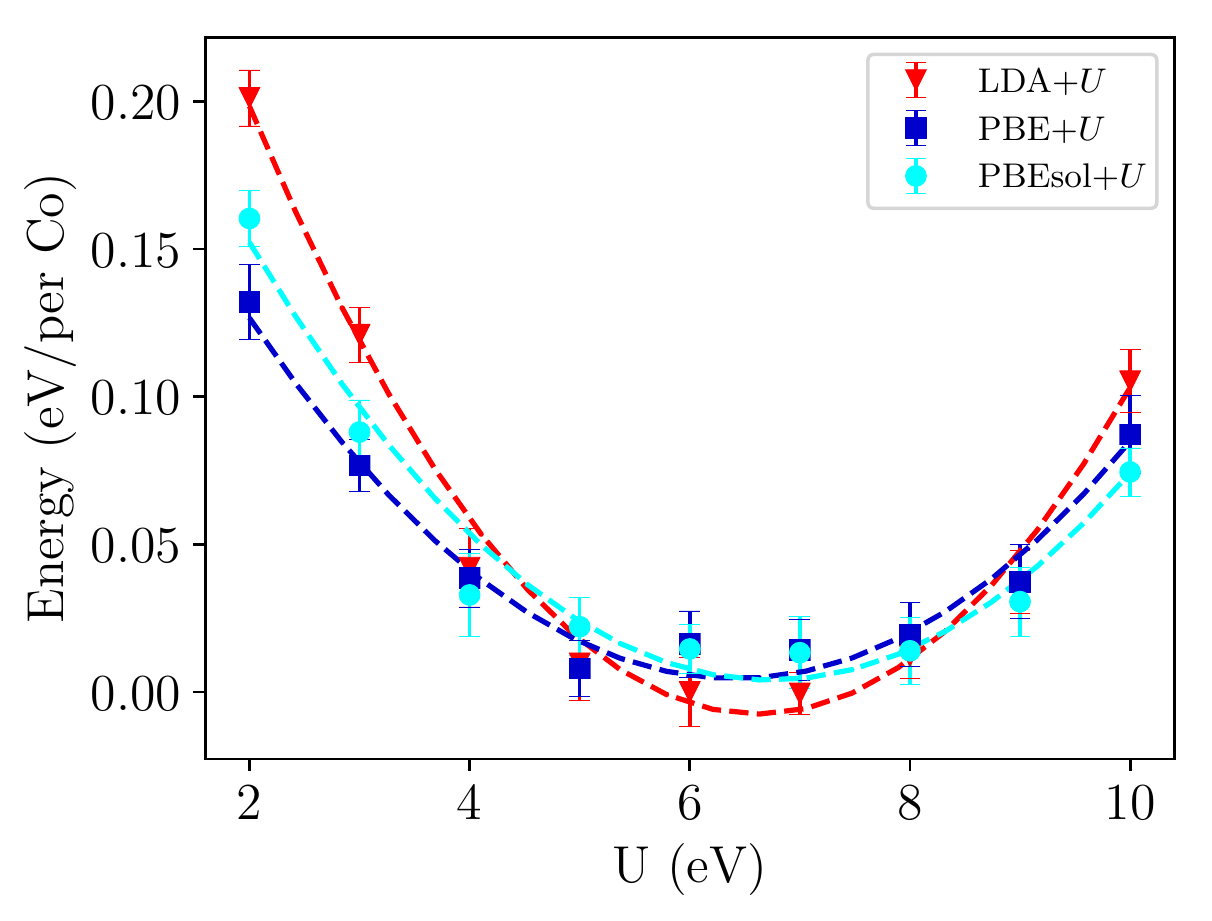}
	\caption{DMC wavefunction optimization of 20-atom HS-AFM bulk LaCoO$_3$ using LDA+$U$, PBE+$U$, and PBEsol+$U$. }
\end{figure}

\begin{figure}[h] \label{fgr:uopt}
	\includegraphics[width=0.5\textwidth]{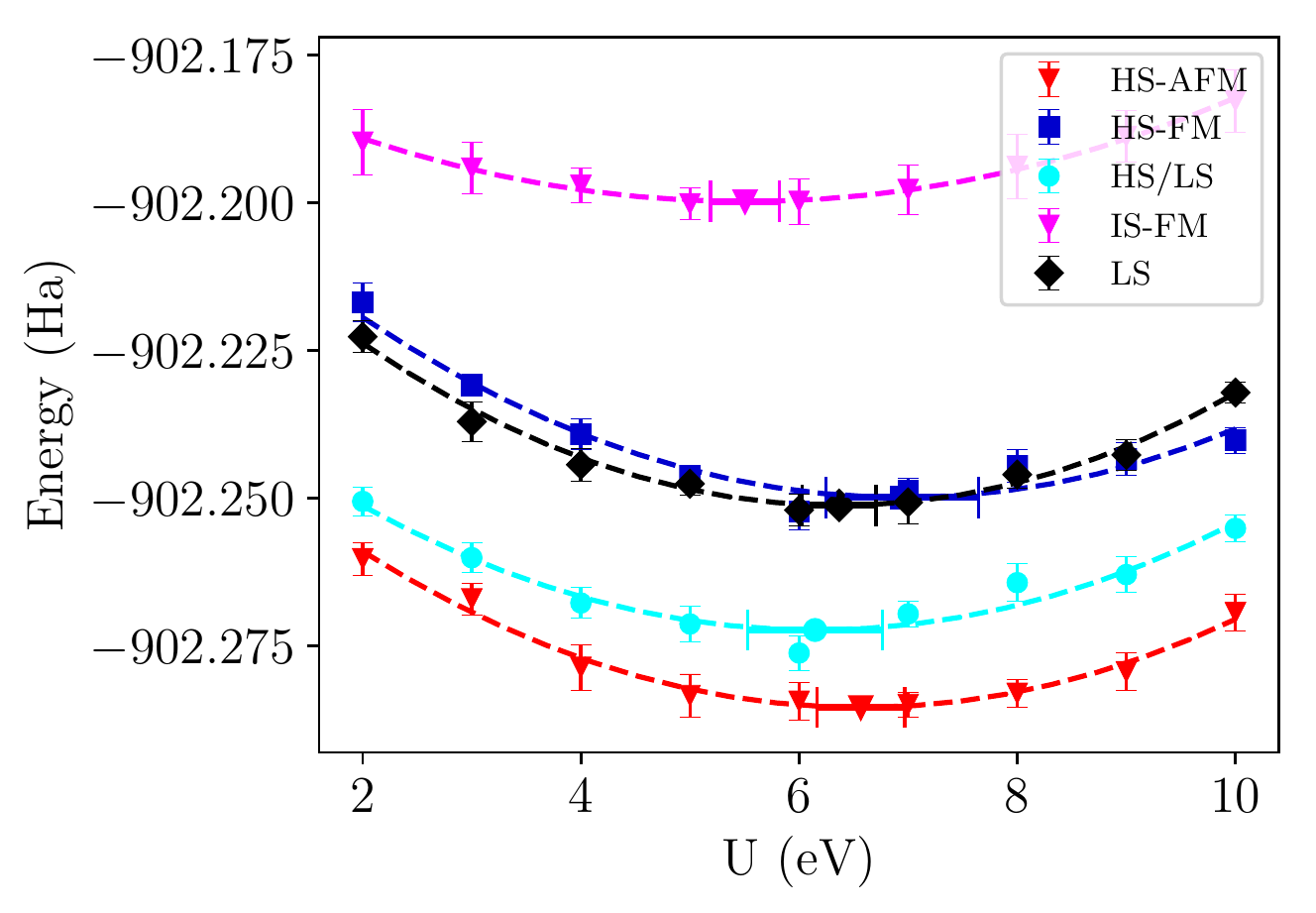}
	\caption{DMC wavefunction optimization of 20-atom bulk LaCoO$_3$ with 2x2x2 reciprocal grid. The wavefunction of each magnetic state is optimized using LDA+$U$. The figure shows that nodal surfaces of the wavefunctions of the all magnetic states of LaCoO$_3$ are optimized near $U=$ 6 eV. }
\end{figure}

\bibliography{citations}

\end{document}